\documentclass{article}

\newif\ificlrstyle
\IfFileExists{iclr2026_conference.sty}{
  \iclrstyletrue
  \usepackage{iclr2026_conference}
    \iclrpreprintcopy 
}{
  \iclrstylefalse

  \usepackage[margin=1in]{geometry}
  \usepackage{times}
}

\usepackage{microtype}
\usepackage{amsmath,amssymb,amsthm}
\usepackage{graphicx}
\graphicspath{{./}}
\usepackage{booktabs}
\usepackage{multirow}
\usepackage{array}
\usepackage{xcolor}
\usepackage{float}
\usepackage{placeins}

\ificlrstyle\else
  \usepackage[numbers]{natbib}
\fi

\usepackage{hyperref}
\hypersetup{hidelinks}

\title{FinBench: Time-Gated Calibration and Uncertainty Benchmarking for Agentic Financial Forecasting}

\author{Rishab Ghosh\\
UT Austin HUMAIN\\
\texttt{rishabghosh@utexas.edu}
\And
Vinay Devarakonda\\
UT Austin\\
\texttt{vinaydev@utexas.edu}}

\begin{document}

\maketitle

\begin{abstract}
Large language models (LLMs) are increasingly used as components of \emph{agentic} systems that observe, plan, and act.
In finance, even ``assistive'' systems become decision-relevant once their outputs are used to size trades or allocate risk.
A key failure mode is the \emph{confidence--competence gap}: a model that is only slightly better than chance but consistently overconfident will, under typical bet-sizing rules, generate negative long-run growth.
Existing benchmarks emphasize semantic understanding or point accuracy, but do not directly test probabilistic calibration under the temporal constraints and non-stationarity that define real markets.

We introduce \textbf{FinBench}, a benchmark designed to evaluate \emph{calibration} and \emph{uncertainty quality} for financial forecasting in a setting that is (i) \emph{strictly time-gated} to avoid look-ahead bias and (ii) evaluated with \emph{strictly proper scoring rules} that penalize hallucinated confidence.
FinBench tasks require models to output (a) a probability of positive return and (b) an 80\% prediction interval for realized log return; evaluation uses the Brier score and the Winkler interval score, along with skill scores against hard baselines.

This paper describes the benchmark specification and reports a small pilot run (one trading day; three liquid tickers; 33 forecasts) as a sanity check of the pipeline.
The pilot illustrates how calibration-sensitive metrics distinguish between ``confident but fragile'' behavior and uncertainty-aware forecasting.
\end{abstract}

\section{Introduction}

LLMs are moving from static question answering to systems that \emph{act}: they can call tools, execute workflows, and make sequential decisions.
In finance, this shift is especially high-stakes because decision quality depends not only on \emph{directional correctness}, but on \emph{how confident the system is} when it takes risk.
If an agent is 55\% accurate but expresses 99\% confidence, standard bet sizing (e.g., Kelly-style sizing) will eventually over-allocate to incorrect views and destroy capital \citep{kelly1956new}.
The failure is not primarily accuracy; it is \emph{miscalibration}.

Most general benchmarks measure semantic understanding or point accuracy.
They do not measure whether the model's stated probabilities reflect true uncertainty, nor do they impose the temporal constraints that matter for finance.
Financial evaluation introduces two structural hazards:
(i) \textbf{non-stationarity}, where regimes change and memorized patterns decay; and
(ii) \textbf{look-ahead bias}, where seemingly innocuous preprocessing can leak future information (e.g., using end-of-day variables to forecast intraday outcomes).
These issues are acute in agentic settings, where a system's errors compound over time.

\textbf{FinBench} is designed as a \emph{calibration and uncertainty benchmark} for agentic finance.
It enforces a strict \emph{time-gating protocol}: when producing a forecast for time $T$, the model is blind to any data (prices, news, derived signals) with timestamps later than $T$.
FinBench uses \emph{strictly proper scoring rules} for both probabilistic classification (Brier score) and interval estimation (Winkler score), so that truthful uncertainty reporting is the optimal strategy \citep{gneiting2007strictly,winkler1972decision,brier1950verification}.

\section{Task Definition}

\subsection{Forecast Targets}

For each ticker and trading day, FinBench issues forecasts at multiple horizons.
In the benchmark design, we emphasize intraday horizons where look-ahead bias is most common; for concreteness we define three example horizons:
\textbf{MORNING} (09{:}30 ET), \textbf{MIDDAY} (12{:}30 ET), and \textbf{CLOSE} (15{:}30 ET).
All forecasts are evaluated against a fixed target time near the end of the day (15{:}55 ET), chosen to be distinct from the forecast horizon.

Let $P_{\mathrm{ref}}$ denote a reference mid-price at the forecast horizon and $P_{\mathrm{target}}$ denote the evaluation mid-price.
The realized log return and the binary direction label are
\begin{equation}
  r = \ln\!\left(\frac{P_{\mathrm{target}}}{P_{\mathrm{ref}}}\right), 
  \qquad
  Y = \mathbb{I}(r > 0).
\end{equation}
Log returns are time-additive and approximately symmetric, making them standard in risk and evaluation settings.

\subsection{Model Output Schema}

Each model must output a structured object containing:
\begin{itemize}
  \item $\hat{p}\in[0,1]$: probability that $Y=1$ (positive return);
  \item $[L,U]$: an 80\% prediction interval for $r$;
  \item an optional natural-language rationale (not used for scoring).
\end{itemize}

\section{Time-Gating Protocol}

The key design goal is to prevent accidental look-ahead.
For a forecast time $T$, the model receives only information with timestamps $\le T$.
This includes both numerical and textual inputs (OHLCV bars, headlines, sentiment features, etc.).
Evaluation is performed at a \emph{fixed} target time.
In our pilot pipeline, we compute $P_{\mathrm{target}}$ using the \emph{last} mid-price observed in the window $[15{:}54{:}30, 15{:}55{:}00]$ ET, and define $P_{\mathrm{ref}}$ analogously at the forecast horizon.
The protocol ensures that ``end-of-day'' quantities are not available at MORNING or MIDDAY horizons.

\section{Metrics}

FinBench prioritizes \emph{calibration} and \emph{interval quality}.

\subsection{Brier Score (Calibration)}

We evaluate probabilistic direction forecasts with the Brier score, a strictly proper scoring rule \citep{gneiting2007strictly,brier1950verification}:
\begin{equation}
  BS = \frac{1}{N}\sum_{i=1}^{N}(\hat{p}_i - Y_i)^2.
\end{equation}
A random guesser with $\hat p=0.5$ yields $BS=0.25$; perfect predictions yield $BS=0$.
To contextualize performance, we report the \emph{Brier Skill Score} (BSS) relative to baselines:
\begin{equation}
  BSS = 1 - \frac{BS_{\mathrm{model}}}{BS_{\mathrm{baseline}}}.
\end{equation}
Positive BSS indicates improvement over the baseline; negative BSS indicates worse-than-baseline calibration.

\subsection{Winkler Score (Interval Quality)}

For an 80\% prediction interval $[L,U]$ for $r$, the Winkler score \citep{winkler1972decision} is
\begin{equation}
  S_{\mathrm{W}} = (U-L) + \frac{2}{\alpha}(L-r)\mathbb{I}(r < L) + \frac{2}{\alpha}(r-U)\mathbb{I}(r > U),
  \qquad \alpha=0.2.
\end{equation}
The score penalizes \emph{cowardice} (wide intervals) and \emph{recklessness} (tight intervals that miss the realized return).

\subsection{Baselines}

To demonstrate genuine value, models should beat simple heuristics:
\begin{itemize}
  \item \textbf{Coin flip}: $\hat p = 0.5$.
  \item \textbf{Base rate}: $\hat p$ equal to a rolling mean of recent outcomes (e.g., 60-day rolling mean of $Y$).
  \item \textbf{Momentum heuristic}: a conditional-probability rule based on recent short-horizon returns (e.g., sign of the last 30 minutes).
\end{itemize}
The pilot CSV includes per-row skill scores against these baselines.

\paragraph{Composite score.}
The pilot CSV also includes a \texttt{composite\_score} column used as an internal aggregation for quick leaderboard views.
Because aggregation choices are design-dependent, we treat Brier/Winkler and skill scores as the primary reported metrics in this paper.

\section{Related Work}

FinBench sits at the intersection of (i) financial-domain language models and benchmarks, (ii) time-series foundation models, and (iii) calibration/safety evaluation.

\paragraph{Financial LLMs and benchmarks.} Domain-specific models such as BloombergGPT and open-source efforts such as FinGPT demonstrate that financial text and market data can be used to adapt LLMs for finance-focused tasks \citep{wu2023bloomberggpt,yang2023fingpt}. Benchmarking efforts like Pixiu provide broad financial evaluation suites, but they are not designed to enforce intraday time-gating for forecasting and do not center strictly proper probabilistic scoring \citep{xie2023pixiu}.

\paragraph{Time-series foundation models.} Recent forecasting models such as Chronos and Moirai, as well as decoder-only foundation approaches for time-series forecasting, target universal numerical forecasting from time-series inputs \citep{ansari2024chronos,woo2024moirai,das2024decoder}. FinBench is complementary: it evaluates models that can integrate \emph{textual context} (news/sentiment) with price history while remaining leak-safe.

\paragraph{Calibration and safety.} Calibration has been shown to be a core failure mode for modern neural networks in high-stakes deployment \citep{guo2017calibration}. For agentic systems, uncertainty misreporting can be safety-critical; broader ML safety agendas emphasize robustness and distribution shift \citep{hendrycks2021unsolved}. FinBench contributes an evaluation harness tailored to agentic finance where miscalibrated confidence is directly penalized.

\paragraph{Agentic systems.} Agentic LLM work highlights how systems can plan and act in interactive environments \citep{park2023generative}. FinBench provides a risk-oriented evaluation substrate for the financial case.

\section{Pilot Experiment}

\subsection{Pilot Setup}

We report a \emph{pilot} run as a pipeline sanity check, using the data the authors recorded in a test run:
\begin{itemize}
  \item \textbf{Date}: 2025-01-15.
  \item \textbf{Tickers}: AAPL, MSFT, NVDA.
  \item \textbf{Horizons}: MORNING (09{:}30), MIDDAY (12{:}30), CLOSE (15{:}30).
  \item \textbf{Target time}: 15{:}55 ET.
  \item \textbf{Models}: GPT-4o, GPT-4o Mini, Llama 3.1 70B, Llama 3.3 70B Instruct, Qwen2.5 72B, DeepSeek-V3 (coverage varies by horizon in this pilot).
  \item \textbf{Total forecasts}: $N=33$ rows in \texttt{data/pilot\_run.csv}.
\end{itemize}
This pilot is \emph{not} intended as a statistically significant leaderboard; it exists to illustrate the benchmark mechanics and confirm that the scoring pipeline produces sensible outputs.

\subsection{Pilot Results}

\begin{figure}[tbp]
  \centering
  \includegraphics[width=0.85\linewidth]{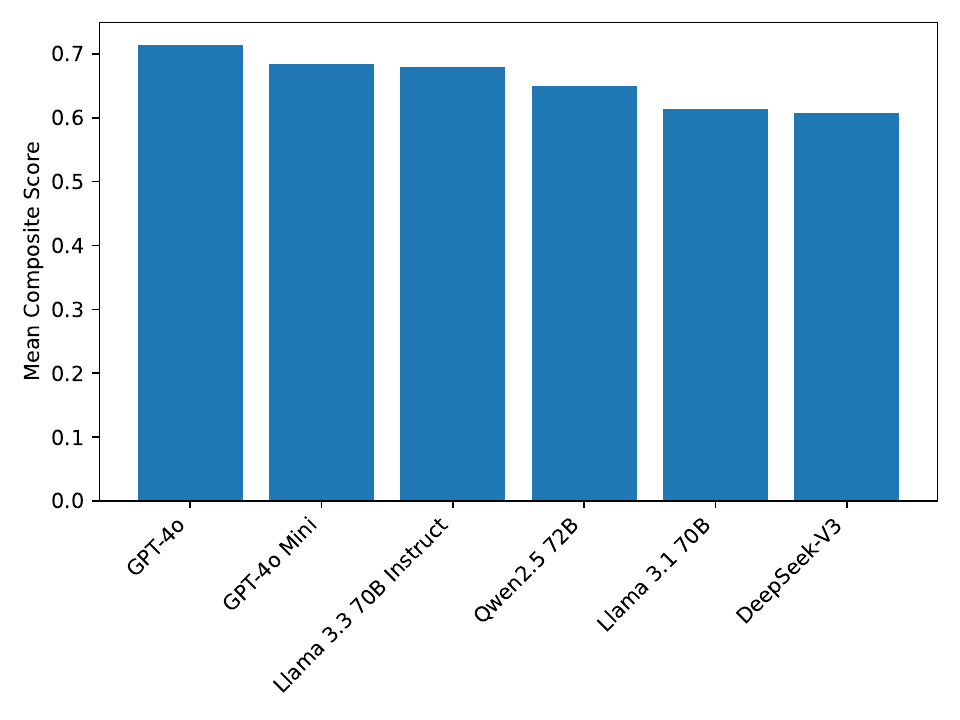}
  \caption{Pilot MORNING-horizon composite scores (mean over three tickers). Higher is better. This is a descriptive visualization only; conclusions require larger samples.}
  \label{fig:pilot_morning}
\end{figure}

The MORNING-horizon results are summarized in \autoref{tab:pilot_morning}, with per-model composite scores shown in \autoref{fig:pilot_morning}.

\begin{table}[t]
\centering
\small
\begin{tabular}{lrrrrrrr}
\toprule
Model & $N$ & Acc. & $BS$ & $BSS_{coin}$ & Cov@80 & Winkler@80 & Comp. \\
\midrule
GPT-4o & 3 & 1.00 & 0.106 & 0.577 & 0.83 & 0.042 & 0.713 \\
GPT-4o Mini & 3 & 1.00 & 0.126 & 0.497 & 0.82 & 0.045 & 0.683 \\
Llama 3.3 70B Instruct & 3 & 1.00 & 0.124 & 0.507 & 0.81 & 0.044 & 0.680 \\
Qwen2.5 72B & 3 & 1.00 & 0.143 & 0.427 & 0.83 & 0.045 & 0.650 \\
Llama 3.1 70B & 3 & 0.67 & 0.179 & 0.283 & 0.83 & 0.046 & 0.613 \\
DeepSeek-V3 & 3 & 0.67 & 0.182 & 0.273 & 0.83 & 0.045 & 0.607 \\
\bottomrule
\end{tabular}
\caption{Pilot (one trading day) results for the MORNING horizon across three liquid tickers (AAPL, MSFT, NVDA). Acc. is directional accuracy from thresholding $\hat p>0.5$. Lower is better for $BS$ and Winkler; higher is better for skill and composite. This table is strictly descriptive and not statistically conclusive.}
\label{tab:pilot_morning}
\end{table}

\paragraph{Key observations from the pilot (descriptive).}
Even at this tiny scale, calibration-sensitive scoring surfaces intuitive failure modes:
(i) all models achieve high MORNING directional accuracy on this day, but their Brier scores and skill separate them (\autoref{tab:pilot_morning});
(ii) two rows (MSFT MORNING) show negative skill vs.\ baselines for Llama 3.1 70B and DeepSeek-V3 due to moderately confident incorrect probabilities (0.55 and 0.56 for an outcome of $Y=0$), yielding Brier scores above 0.30 and negative BSS values in the provided CSV; and
(iii) interval coverage in the pilot is close to the nominal 0.8--0.9 range across rows, but Winkler scores vary with both width and misses, illustrating the ``cowardice vs.\ recklessness'' trade-off.

\begin{table}[t]
\centering
\small
\begin{tabular}{llrrrrrrr}
\toprule
Horizon & Model & $N$ & Acc. & $BS$ & $BSS_{coin}$ & Cov@80 & Winkler@80 & Comp. \\
\midrule
MORNING & GPT-4o & 3 & 1.00 & 0.106 & 0.577 & 0.83 & 0.042 & 0.713 \\
MORNING & GPT-4o Mini & 3 & 1.00 & 0.126 & 0.497 & 0.82 & 0.045 & 0.683 \\
MIDDAY & GPT-4o & 3 & 1.00 & 0.095 & 0.620 & 0.83 & 0.041 & 0.723 \\
MIDDAY & GPT-4o Mini & 3 & 1.00 & 0.112 & 0.550 & 0.82 & 0.044 & 0.697 \\
CLOSE & GPT-4o & 3 & 1.00 & 0.101 & 0.597 & 0.83 & 0.042 & 0.727 \\
CLOSE & GPT-4o Mini & 3 & 1.00 & 0.119 & 0.527 & 0.82 & 0.044 & 0.700 \\
\bottomrule
\end{tabular}
\caption{Pilot comparison of GPT-4o and GPT-4o Mini by horizon (MORNING, MIDDAY, CLOSE).}
\label{tab:pilot_gpt_horizon}
\end{table}

\section{Discussion: Why Calibration Matters for Agentic Finance}

FinBench is motivated by a practical observation: in agentic finance, \emph{uncertainty is the output}.
Two systems with identical directional accuracy can have radically different risk outcomes if one is systematically overconfident.
Proper scoring rules incentivize honest probability reporting, which is a prerequisite for safe decision-making (position sizing, risk limits, human escalation).

\begin{figure}[tbp]
  \centering
  \includegraphics[width=0.85\linewidth]{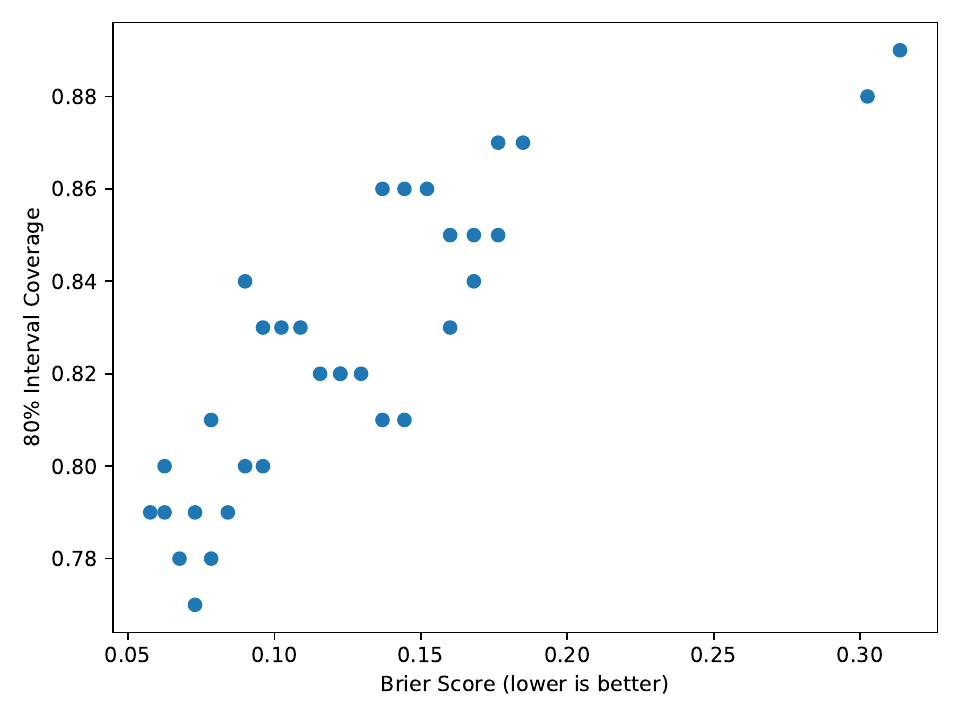}
  \caption{Pilot scatter of Brier score vs.\ 80\% interval coverage across all rows. Calibration and interval quality are distinct; a system can have near-target coverage but poor probability calibration.}
  \label{fig:brier_cov}
\end{figure}

As \autoref{fig:brier_cov} makes clear, calibration and interval quality are distinct axes: a system can land near the target coverage while still reporting poorly calibrated probabilities, so a single accuracy number hides the behavior that matters for risk.

A recurring qualitative pattern we aim to measure at scale is a \emph{rationale--probability mismatch}:
models often generate plausible textual explanations yet output probabilities that do not reflect those explanations.
FinBench isolates this mismatch by scoring only the probabilistic components, while preserving rationales for qualitative analysis.

\section{Limitations and Future Work}

This paper reports a small pilot run and focuses on benchmark specification.
The pilot is too small to draw claims about model superiority, robustness to regime shifts, or profitability.
Future releases of FinBench will:
(i) expand the ticker universe and evaluation horizon across many trading days;
(ii) standardize and publish the exact baseline construction (base-rate and momentum);
(iii) evaluate additional models and prompting strategies under the same time-gating rules; and
(iv) provide stress slices (e.g., macro announcements) to study distribution shift and adversarial fragility.

\section*{Acknowledgments}

We thank Shuozhe Li, Leqi Liu, Tejas Saboo, and Haotian Zhai for helpful discussions and feedback.

\section*{Reproducibility}

FinBench is built to support reproducible and auditable evaluation under strict temporal
constraints. Each evaluation run is defined by an explicit configuration (tickers, horizons,
data windows, baselines, and metric settings), and the benchmark enforces time-gating so that
no input with timestamp greater than the forecast time can enter the context. Model outputs
(probabilities and prediction intervals) and the realized targets used for scoring are logged
along with the metric computations. A reference implementation that regenerates summary tables
and figures from the stored run artifacts will be released alongside an expanded benchmark
release.

\ificlrstyle
  \IfFileExists{iclr2026_conference.bst}{\bibliographystyle{iclr2026_conference}}{\bibliographystyle{plainnat}}
\else
  \bibliographystyle{plainnat}
\fi
\bibliography{references}

\clearpage
\FloatBarrier
\appendix
\section{Pilot Aggregate Table (All Horizons)}

\begin{table}[H]
\centering
\small
\begin{tabular}{lrrrrrrr}
\toprule
Model & $N$ & Acc. & $BS$ & $BSS_{coin}$ & Cov@80 & Winkler@80 & Comp. \\
\midrule
GPT-4o & 9 & 1.00 & 0.100 & 0.598 & 0.83 & 0.042 & 0.721 \\
GPT-4o Mini & 9 & 1.00 & 0.119 & 0.524 & 0.82 & 0.044 & 0.693 \\
Llama 3.3 70B Instruct & 3 & 1.00 & 0.124 & 0.507 & 0.81 & 0.044 & 0.680 \\
Qwen2.5 72B & 3 & 1.00 & 0.143 & 0.427 & 0.83 & 0.045 & 0.650 \\
Llama 3.1 70B & 6 & 0.83 & 0.152 & 0.390 & 0.83 & 0.045 & 0.647 \\
DeepSeek-V3 & 3 & 0.67 & 0.182 & 0.273 & 0.83 & 0.045 & 0.607 \\
\bottomrule
\end{tabular}
\caption{Pilot aggregate across all horizons in the pilot CSV. Note that not all models are evaluated at all horizons in this pilot; $N$ therefore varies by model. Interpret only as a sanity check that the pipeline produces sensible metrics.}
\label{tab:pilot_overall}
\end{table}

\section{Notation Quick Reference}

\begin{table}[H]
\centering
\small
\begin{tabular}{ll}
\toprule
Symbol & Meaning \\
\midrule
$P_{\mathrm{ref}}$ & Reference mid-price at the forecast horizon. \\
$P_{\mathrm{target}}$ & Evaluation mid-price at 15{:}55 ET (or last mid-price in a short window ending at 15{:}55). \\
$P_{\mathrm{mid}}$ & Mid-price computed from high/low: $P_{\mathrm{mid}}=(P_{\mathrm{high}}+P_{\mathrm{low}})/2$. \\
$r$ & Realized log return, $r=\ln(P_{\mathrm{target}}/P_{\mathrm{ref}})$. \\
$Y$ & Direction label, $Y=\mathbb{I}(r>0)$. \\
$\hat{p}$ & Model-reported probability that $Y=1$. \\
$[L,U]$ & Model-reported 80\% prediction interval for $r$. \\
$BS$ & Brier score for probabilistic direction forecasts. \\
$BSS$ & Brier skill score vs.\ a baseline (coin-flip, base-rate, momentum). \\
$S_{\mathrm{W}}$ & Winkler interval score for $[L,U]$ with $\alpha=0.2$. \\
\bottomrule
\end{tabular}
\caption{Notation used throughout the FinBench task and metrics definitions.}
\label{tab:notation}
\end{table}

\begin{figure}[tbp]
  \centering
  \includegraphics[width=0.85\linewidth]{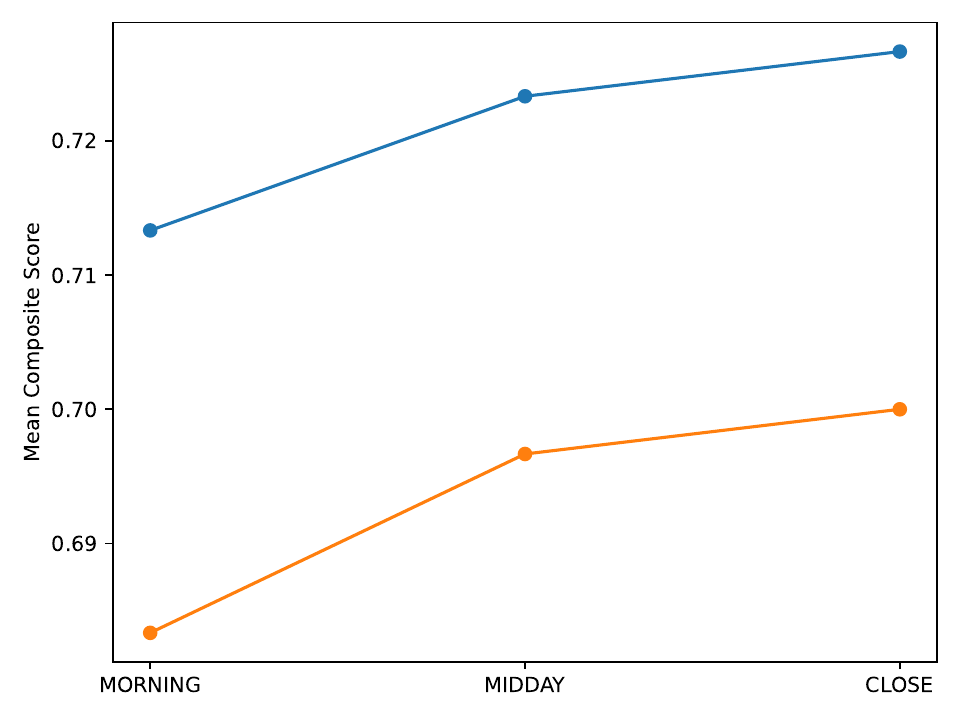}
  \caption{Pilot comparison of GPT-4o vs.\ GPT-4o Mini across horizons (mean composite). In this pilot, both models appear stable across horizons, but results are descriptive only.}
  \label{fig:pilot_horizon}
\end{figure}

\end{document}